\documentclass[conference]{IEEEtran}

\ifCLASSINFOpdf
\else
\fi

\usepackage{comment, xcolor, listings, graphicx, caption, amsmath, algorithm, algorithmic, mathtools, cleveref, booktabs, mdframed, multirow, array}
\usepackage[frozencache,cachedir=minted]{minted}
\usepackage[normalem]{ulem}

\crefformat{section}{\S#2#1#3} %
\crefformat{subsection}{\S#2#1#3}
\crefformat{subsubsection}{\S#2#1#3}

\AtBeginEnvironment{minted}{\dontdofcolorbox}
\def\dontdofcolorbox{\renewcommand\fcolorbox[4][]{##4}}

\newcommand{\sysname}{{\sc DisasLLM}}
\newcommand{\todoc}[2]{{\textcolor{#1}{\textbf{#2}}}}
\newcommand{\todored}[1]{{\todoc{red}{\textbf{[#1]}}}}

\newcommand{\todo}[1]{\todored{TODO: #1}}

\newcommand{\ronghua}[2]{{\textcolor{red}{\sout{#1}#2}}}
\newcommand{\hang}[1]{\textcolor{blue}{[Hang: #1]}}
\newcommand{\ignore}[1]{}

\begin{document}
\title{Disassembling Obfuscated Executables with LLM}

\author{
    \IEEEauthorblockN{
		Huanyao Rong\IEEEauthorrefmark{1},
		Yue Duan\IEEEauthorrefmark{2},
		Hang Zhang\IEEEauthorrefmark{1},
		XiaoFeng Wang\IEEEauthorrefmark{1},
		Hongbo Chen\IEEEauthorrefmark{1},
		Shengchen Duan\IEEEauthorrefmark{2},
		Shen Wang\IEEEauthorrefmark{2}}
    \IEEEauthorblockA{\IEEEauthorrefmark{1}Indiana University Bloomington}
    \IEEEauthorblockA{\IEEEauthorrefmark{2}Singapore Management University}
}

\maketitle

\begin{abstract}

Disassembly is a challenging task, particularly for obfuscated executables containing junk bytes, which is designed to induce disassembly errors. Existing solutions rely on heuristics or leverage machine learning techniques, but only achieve limited successes. Fundamentally, such obfuscation cannot be defeated without in-depth understanding of the binary executable's semantics, which is made possible by the emergence of large language models (LLMs).
In this paper, we present \sysname{}, a novel LLM-driven dissembler to overcome the challenge in analyzing obfuscated executables. \sysname{} consists of two components: an LLM-based classifier that determines whether an instruction in an assembly code snippet is correctly decoded, and a disassembly strategy that leverages this model to disassemble obfuscated executables end-to-end.
We evaluated \sysname{} on a set of heavily obfuscated executables, which is shown to significantly outperform other state-of-the-art disassembly solutions.

\end{abstract}

\section{Introduction}

Disassembly, a process of translating machine-readable bytes into human-readable assembly instructions, is a critical step of binary reverse engineering, enabling important downstream tasks such as decompilation~\cite{Cifuentes1995DecompilationOB, Balakrishnan2007DIVINEDV}, binary rewriting~\cite{FloresMontoya2019DatalogD, Kim2023ReassemblyIH, Wang2016UROBOROSIS} and vulnerability detection~\cite{Shoshitaishvili2016SOKO, bai2020idea}.
One major challenge in disassembly is correct identification of instruction boundaries within a byte sequence: due to the mixed storage of program code and data, or variable-length instructions such as those of x86, the locations where instructions start are often hard to determine, leading to disassembly errors not only in the instructions at the boundaries but also in the follow-up ones.
This challenge has been exploited by various obfuscation techniques, such as opaque predicate, call stack tampering, signal-based obfuscation~\cite{Popov2007BinaryOU}, to confuse a disassembler, which utilize junk bytes to increase the difficulty in delimiting binary sequences.
To address this problem, existing techniques leverage symbolic execution~\cite{Bardin2017BackwardBoundedDT, Shoshitaishvili2016SOKO}, heuristics~\cite{Pang2020SoKAY, FloresMontoya2019DatalogD, Miller2019ProbabilisticD, Krgel2004StaticDO, Tung2020AHA}, or machine learning~\cite{Pei2020XDAAR, Yu2022DeepDiLA, Krishnamoorthy2009StaticDO}. However, their effectiveness has been found to be rather limited.

\subsection{Limitations of Existing Work} \label{CurrentLimitations}

Symbolic execution-based approaches~\cite{Shoshitaishvili2016SOKO} remove junk binaries within infeasible paths of a program to help determine instruction start addresses through analyzing the program's control and data flows.  Their applications to binary deobfuscation, however, are hindered by
path- and state-explosions
as well as the presence of hard-to-solve path conditions -- weaknesses exploited by adversarial techniques~\cite{Xu2018ManufacturingRB}.
A variation of the approaches, concolic execution~\cite{Bardin2017BackwardBoundedDT},  analyzes a single execution trace of a binary executable instead of its whole program,
and therefore is more scalable, but only covers the portion of the binary sequence related to the execution trace.

Some prior research utilizes manually constructed, domain specific heuristics to decide the correct instruction boundaries~\cite{Pang2020SoKAY, FloresMontoya2019DatalogD, Miller2019ProbabilisticD, Krgel2004StaticDO, Tung2020AHA}.
These heuristics are usually based upon shallow syntactical features, such as unusual opcode of the instructions recovered from wrong start addresses.
Such approaches tend to be less robust since they rely heavily on empirical observations and assumptions,
which could be easily broken, either by circumvention or due to the situations they cannot fit into.

Recently, machine learning (ML) has shown promise in addressing this problem.
An ML-based approach automatically learns the features of valid and invalid instruction sequences to identify correctly recovered instructions.
For this purpose, an early work~\cite{Krishnamoorthy2009StaticDO} relies on feature engineering and decision trees. These techniques, however, are found to be inadequate for comprehensive and accurate detection of instruction boundaries.
A more recent solution -- XDA~\cite{Pei2020XDAAR} trains a transformer~\cite{Vaswani2017AttentionIA} \ignore{on raw bytes }to analyze individual bytes, predicting whether each of them is the start of an instruction. However, since these bytes are not decoded first, the model must learn the decoding by interpreting the raw bytes directly. This limits the model's capacity to learn the semantic meaning of the machine code, as it first has to learn the decoding process.
Another state-of-the-art work is DeepDi~\cite{Yu2022DeepDiLA}, which employs superset disassembly~\cite{Bauman2018SupersetDS} to decode all possible addresses, before instruction validity can be determined with a graph neural network (GNN) on the instruction flow graph.
However, due to the limited semantic and contextual learning capability of GNN, it has been demonstrated that DeepDi is vulnerable to dedicated adversary obfuscation techniques, which drastically drop its F1 score from 91.64\% to 16.67\%~\cite{Zhang2023PELICANEB}.
A primary reason for their limitation is their restricted model size, which further limits their capability to understand the complex semantics of decoded assembly instructions. One might suggest increasing the model size and the training dataset; however, expanding their dataset to trillions of bytes and their parameter size to billions, similar to current state-of-the-art large language models, is simply impractical. Therefore, a transfer learning approach that fine-tunes the current large language model is a more promising solution by comparison.
Finally, all aforementioned ML methods completely exclude the more traditional but widely-used and time-tested disassembly approaches (e.g., recursive disassembling algorithm),
which unnecessarily lose some benefits.
For example, traditional disassembly provides the insight that if one instruction is valid, its following instructions are also likely to be valid, while any overlapping instructions are likely to be invalid. However, these methods do not take advantage of this insight.

\subsection{Our Method}

Our observation is that an experienced human expert usually can effectively recognize instruction boundaries: disassembled instructions at incorrect offsets tend to deviate significantly from their contexts,
either at the syntax level or at the semantic level, and can therefore be captured by human experts.
Existing approaches, whether heuristic-based or ML-powered, are not designed to consider semantic information in valid instruction identification and have limited capability in doing so.
The recent advancement of generative large language models (LLMs), however, brings unprecedented opportunities to find a much more effective solution, as LLMs are meant to predict missing tokens from their contexts, and have demonstrated a remarkable
comprehension ability closer to that of human analysts than all existing techniques~\cite{Vaswani2017AttentionIA, Devlin2019BERTPO, Achiam2023GPT4TR, Touvron2023Llama2O, Chen2021EvaluatingLL, Rozire2023CodeLO}.
One straightforward solution is thus to let LLM inspect the whole byte sequence and decide the instruction boundaries (similar to XDA~\cite{Pei2020XDAAR}),
or to enumerate all possible disassembly results with different offsets and then rely on LLM to pick the correct ones (similar to DeepDi~\cite{Yu2022DeepDiLA}).

However, such a direct application of the LLM does not fully leverage its capabilities for the following reasons.
(1) Presenting machine code in bytes or in superset disassembly does not convey their semantic information in the most effective form: the LLM may struggle to understand the raw bytes directly, or may become overwhelmed by the redundant information in superset disassembly.
(2) Such a naive approach can be quite expensive: it requires inspecting many potential addresses, resulting in a large number of queries to an LLM, which incurs significant delays or monetary costs.
(3) Current decoder LLMs are designed for text generation tasks. However, the task of identifying instruction validity is more akin to a classification task, which is generally handled by encoder models.
(4) Current LLMs are not trained on a large corpus of valid and invalid assembly text, which limits their direct capability to identify such validity.

To address these problems,
our insight is that disassembly errors due to incorrect instruction boundaries are not prevalent (e.g., due to the performance cost, usually only some important code snippets will be obfuscated).
In a typical reverse engineering session, a traditional disassembler (though not resilient to obfuscation) can already recover most valid instructions,
while human analysts only need to fix errors in limited regions.
Therefore, our idea is to integrate the LLM into a traditional disassembler -
such a hybrid system mimics the human expert's reverse engineering process automatically and efficiently.
Based on this idea, we propose \sysname{},
a systematic disassembly approach resilient to obfuscation, built upon an LLM-based classifier capable of identifying the validity of decoded instructions.
\sysname{} consists of two components: an LLM-based validity classifier that determines whether instructions are correctly decoded and a disassembly strategy that leverages this classifier to disassemble obfuscated executables end-to-end efficiently and effectively.

\subsubsection{LLM-based Validity Classifier}

To train this model, we fine-tuned Llama 3 8B~\cite{llama3modelcard} using the method proposed by LLM2Vec~\cite{BehnamGhader2024LLM2VecLL}, which converts a decoder model into an encoder model for NLP understanding tasks, including the token classification we require. This method allows us to address the problem (3) mentioned previously.

To prepare the unsupervised text dataset for the fine-tuning, we first compiled source functions from AnghaBench~\cite{Silva2021ANGHABENCHAS} and collected all instruction boundary addresses as the ground truth. We then disassembled these compiled binaries according to the ground truth into assembly text, including validly decoded instructions and invalid ones in the comments. This allows the model to learn the patterns of both valid and invalid instructions. Additionally, to further train a classifier, we gather the supervised dataset by disassembling the compiled AnghaBench binaries using our disassembly strategy (introduced soon). However, we replace the classifier that should result from this training task with a classifier based on the ground truth of the instruction addresses, ensuring the classifier's results are always correct. The dataset is then obtained by recording all outcomes of these classifications. These training process allows us to solve problem (4).

\subsubsection{Our Disassembly Strategy}

Given an obfuscated binary executable, \sysname{} disassembles its code memory region (i.e., the \texttt{.text} section) using a combination of linear disassembly and recursive disassembly to obtain an initial result.
Based on this result, \sysname{} then verifies the validity of each decoded instruction from the initial disassembly and tries to fix the incorrect disassembly if applicable. To do this both efficiently and effectively, instead of checking all decoded instructions simultaneously or each instruction individually, \sysname{} uses a pre-filter to examine $N$ adjacent instructions at once to confirm the validity of instructions with very high or low probability values. Based on the pre-filter results, \sysname{} then checks the remaining instructions one by one. Such optimized design aims to address the problem (2) mentioned previously.

To best represent the semantic information for an LLM to check validity, in order to verify the validity of one or more instructions, \sysname{} extracts all related instructions in the disassembly result with a breadth-first search algorithm, such as those with reference relations.
Additionally, since instruction fixing and checking are conducted interchangeably, related instructions that are already fixed can aid in identification. The method addresses the problem (1) of directly applying LLM.

\subsubsection{Evaluation Results}

We evaluated \sysname{} on a set of obfuscated executables with heavily inserted junk bytes. Compared to the state-of-the-art approach, \sysname{} achieved an approximate 40\% improvement in detecting the first instruction after the junk byte region.
We also present the relevant metrics of our classifier model, both when used in conjunction with the disassembly strategy and when used independently, demonstrating its effectiveness in identifying valid instructions.

\subsubsection{Contributions}

We have proposed \sysname{} and made the following contributions to the field of disassembly:

\begin{itemize}
\item We trained an LLM-based classifier model to identify the validity of instructions within a snippet of decoded assembly code by treating it as a token classification problem.
\item Building upon the validity identification capability of the model, we propose a disassembly strategy that leverages the model to disassemble obfuscated binary executables.
\item We evaluated \sysname{} on a benchmark based on OLLVM containing a set of obfuscated executables. The results show that it significantly outperforms other disassembler counterparts.
\end{itemize}

\section{Background}

\subsection{Binary Disassembly}
\label{bg-disas}

Binary disassembly translates the raw bytes into human-readable assembly instructions, serving as the vital first step of binary reverse engineering. We discuss some common disassembly techniques.

\vspace{2pt}\noindent\textbf{Linear and recursive disassembly}.
Linear disassembly sequentially processes each instruction.
After decoding one instruction, the disassembling continues from the next byte which is treated as the start of the subsequent instruction.
This strategy exploits the continuity of compiler code generation.
Recursive disassembly, on the other hand, processes the instructions based on observable control flows.
For example, after decoding a jump instruction with a constant target address,
the disassembler assumes that the target address starts a new instruction and continues disassembling from there.
These two strategies could be combined to achieve the optimal results~\cite{Pang2020SoKAY},
where the code ``holes'' not disassembled by the recursive disassembly are handled by linear disassembly (e.g., used by Angr~\cite{Shoshitaishvili2016SOKO}).

However, these traditional approaches can be problematic when data or junk bytes present (e.g., at the next byte or the jump target address), whose decoding leads to invalid instructions.
Even worse, the addresses of invalid instructions can overlap with the valid ones, causing the latter to be disregarded.
Nonetheless, one interesting phenomenon of this traditional disassembly process is that even if one invalid starting address is disassembled, a correct starting address will eventually be encountered after several wrongly decoded instructions
(e.g., imagine a 7-byte sequence containing 1 leading junk byte and 2 subsequent valid instructions, while it may be wrongly disassembled into 3 invalid instructions, starting from the 8th byte the disassembly will return to the normal).
This phenomenon is referred to as the ``self-repairing''~\cite{Linn2003ObfuscationOE}, increasing the disassembly stability but also making the disassembly errors more stealthy.

\vspace{2pt}\noindent\textbf{Disassembly based-on valid instruction identification}.
To overcome the weakness of traditional disassembly, learning-based or heuristic-based strategies are proposed to decide the validity of the disassembled instruction sequences and the correct instruction boundaries, based on specific features. 
However, as described in \cref{CurrentLimitations}, current approaches show limited effectiveness in defeating obfuscation.

\subsection{Large Language Model}

Based on vast amounts of training data,
the large language model (LLM)~\cite{Vaswani2017AttentionIA} marks a significant advancement in natural language processing (NLP) and program code understanding~\cite{Rozire2023CodeLO, Chen2021EvaluatingLL}.
LLMs can be categorized into encoder models and decoder models. Encoder models primarily focus on understanding the text, while decoder models are mainly designed for text generation.

Token classification~\cite{TokenClassification}, which assigns labels to tokens in a text, is one task of encoder models.
For example, named entity recognition labels a text entity with predefined categories such as person names, organizations, locations, and dates.
This labeling takes into account both the entity and its contexts.

A recent work, LLM2Vec~\cite{BehnamGhader2024LLM2VecLL}, shows that a decoder model can be converted into an encoder model through fine-tuning.
The technique starts with enabling bidirectional attention of the transformer so that the preceding tokens can attend to the following tokens in the self-attention layer. However, since the generative decoder model is not trained with such bidirectional attention, the effectiveness may drop for such change.
To address this, the LLM is fine-tuned in an unsupervised manner using a corpus of texts as the dataset, just like the pre-training step, but instead of training to predict the next token given the preceding tokens, LLM2Vec leverages the technique known as masked next token prediction (MNTP), which trains to predict the randomly masked tokens given the unmasked ones in a training token sequence, a training approach commonly used by encoder model~\cite{Devlin2019BERTPO}. With the fine-tuned model, the embedding vector of one or more tokens under the text context can be computed, which can then be used to train another classifier model for the token classification task.
The encoder model enables more effective use in token classification tasks, as demonstrated in their results~\cite{BehnamGhader2024LLM2VecLL}. In the following sections, we will describe how to utilize this technique to construct our validity classifier model.

\section{Motivation} \label{Motivation}

In this section, we exemplify the limitations of state-of-the-art disassemblers in detail and demonstrate how LLMs can help, which motivates our work.

\subsection{Limitations of Existing Work} \label{CurrentChallenge}

DeepDi~\cite{Yu2022DeepDiLA} is the state-of-the-art disassembler resilient to obfuscation due to its superset disassembling and GNN-based valid instruction sequence identification.
However, obfuscation designed to thwart disassembly can still drastically reduce its effectiveness. 
To show this, we compile a simple C program containing all C syntactic features and obfuscate the resulting binary by (1) flattening the control flow~\cite{Lszl2009OBFUSCATINGCP} with OLLVM~\cite{Junod2015ObfuscatorLLVMS},
and (2) inserting the junk bytes between basic blocks.
This obfuscation puts junk bytes in unreachable code regions, testing the disassembler's instruction boundary identification ability.
DeepDi's performance on this binary is summarized in Table \ref{DeepDiStudy}.
We define the false negative (FN) as ``valid instructions not shown in the disassembly result'',
while false positive (FP) as ``invalid instructions overlapping with the valid ones shown in the disassembly result'',
as non-overlapping invalid instructions do not affect the disassembly of valid ones.
Precision, recall, and the F1 score are defined accordingly.
The ``All'' row considers all instructions in the binary when computing the above metrics,
while the ``Junk'' row only includes the first valid instructions following the inserted junk bytes in the statistical scope,
because they are most heavily affected by the junk bytes,
how well they can be disassembled directly reflects the disassembler's anti-obfuscation ability.
As seen in Table \ref{DeepDiStudy},
though DeepDi achieves good metrics in the ``All'' row, its performance significantly reduces in the ``Junk'' row.
This suggests that DeepDi can successfully disassemble most valid instructions whose boundaries are not impacted by the junk bytes, but falls short when handling the obfuscated regions.

\begin{table}[]
\centering
\caption{DeepDi on the small obfuscated binary.}
\label{DeepDiStudy}
\footnotesize
\begin{tabular}{@{}cccc@{}}

\toprule
 & Precision  & Recall & F1  \\
\midrule
All & 0.94 & 0.92 & 0.93 \\ \hline
Junk & 0.57 & 0.47 & 0.52 \\
\bottomrule
\multicolumn{4}{l}{Precision = TP / (TP + FP), Recall = TP / (TP + FN)} \\
\multicolumn{4}{l}{F1 = 2 * Precision * Recall / (Precision + Recall)} \\
\end{tabular}
\end{table}

\subsection{LLM's Potential}

A key observation is that invalid instructions decoded at the wrong offsets exhibit a very different appearance compared to valid instructions within the context.
Thus, an experienced human reverse engineer can usually discern invalid instructions from valid ones easily.
Thanks to the self-repairing feature (\cref{bg-disas}), even with junk bytes,
traditional disassembly algorithms can still recover many valid instructions, albeit mixed with invalid ones.
Therefore, in principle, a human expert can inspect these disassembly results to identify and correct the invalid instructions. However, this process requires substantial manual effort and is time-consuming, especially for large binaries.

Our intuition is to replace human experts with an LLM, given its outstanding capabilities of comprehending semantic information in both natural languages~\cite{Achiam2023GPT4TR} and programming languages~\cite{Chen2021EvaluatingLL},
including assembly code~\cite{Jiang2023NovaGL}.
The advancements of LLMs provide a unique opportunity to develop a system resembling the human-involved disassembling process,
while fully automatic and efficient.

\section{Problem Statement}

\sysname{} aims to disassemble executables obfuscated to thwart the disassembly process. For example, junk bytes are inserted in the code region of such an executable, causing the linear disassembly approach to decode invalid instructions that may even overlap with valid ones.
Additionally, opaque predicates can cause a conditional jump, which would never actually occur, leading to the recursive disassembly to decode the unreachable branch target and potentially overlapping with valid instructions. To sum up, \sysname{} is robust enough to handle almost all types of obfuscation designed to defeat disassembly, such as opaque predicate, call stack tampering, signal-based obfuscation~\cite{Popov2007BinaryOU}, control flow flattening~\cite{Lszl2009OBFUSCATINGCP}, that are combined with junk byte insertion.

However, \sysname{} does not aim to address self-modifying code or packing. In these scenarios, an analyst should first apply other approaches~\cite{Royal2006PolyUnpackAT, Bonfante2015CoDisasmMS} to extract the recovered code bytes before using our system. Additionally, \sysname{} is focused solely on disassembling the executable rather than other downstream deobfuscation tasks (e.g., recovering the original program from VM-based obfuscation~\cite{Yadegari2015AGA}).

\section{The Disassembly Strategy} \label{Disassembly}

\begin{figure}
\center
\includegraphics[scale=0.85]{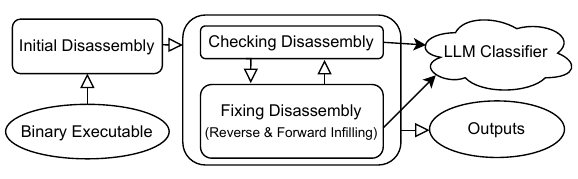}
\captionsetup{skip=5pt, name={Figure}}
\caption{Workflow of \sysname{}. The hollow arrow denotes the sequential procedure. The solid arrow represents the invocation with information returned.}
\label{Workflow}
\end{figure}

Figure~\ref{Workflow} illustrates the overview of \sysname{}. Given a binary executable, \sysname{} first performs an initial disassembly on its code region (\cref{InitialDisasm}). Then, our system checks the validity of each decoded instruction (\cref{CheckDisasm}), during which it also attempts to fix disassembly errors when certain conditions are met (\cref{FixDisasm}).
Both procedures invoke the LLM-based classifier to check the validity of one or more instructions by representing the decoded instructions as text (\cref{RepresentInst}). To enhance the efficiency of using LLM, \sysname{} employs a batching mechanism to exploit the parallel capabilities of GPUs (\cref{Batching}).
Once all decoded instructions are checked and corrected, \sysname{} outputs the final disassembly results.

\subsection{Initial Disassembly} \label{InitialDisasm}

Given the raw bytes of the code region of a binary executable, \sysname{} disassembles it using a combination of linear disassembly and recursive disassembly. The algorithm is essentially the same as recursive disassembly, but for each branch instruction, such as \texttt{jmp} and \texttt{ret}, the next instruction starting at its end address is also considered a candidate. Additionally, we employ a more aggressive recursive approach that tracks immediate numbers appearing in non-branch instructions, similar to \cite{FloresMontoya2019DatalogD}.

The main reason for relying on the traditional approach instead of the more aggressive superset disassembly~\cite{Bauman2018SupersetDS}, which decodes every possible address, is two-fold. First, our linear plus recursive disassembly approach can produce sufficient results as a starting point more efficiently than super disassembly. This is due to the self-repairing feature of the traditional disassembly approach - even when an executable is heavily obfuscated with junk bytes inserted in the code region, our disassembly result still contains many validly decoded instructions. Second, as demonstrated later in Section \cref{EvalModel}, the relevant instructions from traditional disassembly are sufficient to identify an instruction's validity correctly. In contrast, superset disassembly provides excessive redundant information that could interfere with performance.

After the initial disassembly, \sysname{} constructs a graph where the vertices represent basic blocks (sequences of decoded instructions), and the edges represent references between these blocks. We refer to this graph as the \textit{disassembly graph}.
The references include not only control flow relations but also references caused by immediate numbers due to a more aggressive recursive disassembly algorithm. Additionally, since \sysname{} also incorporates linear disassembly, the block resulting from the continued decoding of a non-conditional branch instruction creates a special edge. This edge connects the source block containing the branch instruction to the block starting at the end address of the branch instruction.
Only the first instruction of each block can be referenced. Therefore, during the disassembly process, if an instruction in the middle of a block is referenced, the block must be split into two blocks, with an edge connecting them. Such edges allow the breadth-first search algorithm to find related instructions when constructing the snippet for the classification.

\subsection{Checking Disassembly Results} \label{CheckDisasm}

With the disassembly graph obtained from \cref{InitialDisasm}, \sysname{} performs overlap minimization (described later in \cref{HandleOverlap}). Following this, \sysname{} verifies the validity of each decoded instruction. Given a piece of decoded assembly instructions in text form, our task is to classify the validity of one or more instructions within this snippet. We consider this task a classical token classification problem, such as named entity recognition, which aims to classify which entity each word in a sentence belongs to. These tasks are highly similar in the sense that we can consider the instruction as the word and the validity as the entity.

We can consider two naive approaches to classify each instruction disassembled from \cref{InitialDisasm}. 1) We can represent all blocks in the entire disassembly graph as texts for the classification of each instruction. However, for large binaries consisting of thousands of instructions, the text will exceed the maximum token length supported by the model. 2) Instead of classifying all instructions individually, we can classify each instruction individually. This approach, in principle, can be more effective than the previous one because identifying a subsequent instruction can leverage the preceding validity results of other instructions, providing richer contextual information. However, this method can be too slow for large binaries.

Therefore, we propose a trade-off between these two methods.
Firstly, \sysname{} checks $N=16$ multiple adjacent instructions simultaneously to conduct a prefilter on the disassembled instructions. We obtain the probability of each instruction being valid by obtaining the sigmoid~\cite{Han1995TheIO} value from the classifier. In the prefilter stage, \sysname{} empirically sets the validity of instructions with a probability $p > 0.95$ as valid and $p < 0.05$ as invalid.
According to our evaluation in \cref{IntermMetrics}, these instructions already constitute most of the total number of instructions in the disassembly graph. Additionally, \sysname{} deletes the blocks in which all instructions are invalid, removing most invalid overlapped blocks. After block deletion, overlap minimization is further performed to reduce the overlapped regions in the new disassembly graph.
After the prefilter stage, we check the remaining instructions one by one. This allows \sysname{} to exploit the validity information of any related instructions classified during the prefilter stage.

To examine the validity of instructions, presenting only the instructions themselves is insufficient; their related instructions are also required as input to provide contextual information. These related instructions are obtained by conducting a breadth-first search starting from the instructions to be examined.
To create the graph for the BFS task, we consider the disassembly graph to be a graph with instructions as vertices. In this graph, an instruction within a block is connected to the next instruction in the block, and the edges that connect blocks are now used to connect the instructions.
When performing BFS, \sysname{} traverses both forward and reverse edges, ensuring that given an instruction $I$, both the instructions referenced by $I$ and the instructions that reference $I$ are traversed.
The BFS stops when the number of related instructions reaches a limit, which we have empirically set to $32$, as this should provide sufficient contextual information for identifying validity.
\sysname{} then extracts the blocks containing instructions visited by BFS. These blocks can either be the original blocks from the disassembly graph if all instructions within the blocks are visited or split blocks if only part of the instructions are visited. These blocks are represented as texts for token classification to check validity.

\subsection{Fixing Disassembly Results} \label{FixDisasm}

Note that not all authentic instructions are decoded in the disassembly graph; instead, these instructions can overlap with invalid ones in the graph. Therefore, \sysname{} must correct the disassembly results obtained from \cref{InitialDisasm} rather than merely classifying the validity of each instruction.

To achieve this, during the validity checking described above, \sysname{} keeps track of the validity of the memory regions.
A valid region is defined as a region occupied by instructions already identified as valid. In contrast, an invalid region is a region that does not contain any parts of valid or unidentified instructions (i.e., it only contains invalid instructions). Other regions are considered unidentified.
When an invalid region is produced between two valid regions, \sysname{} attempts to find the correct disassembly within the invalid region. All invalid instructions within this region are removed from the disassembly graph.
\sysname{} first attempts reverse infilling to find the valid instruction sequence that ends at the start address of the valid region. Next, \sysname{} conducts forward infilling to find any missing blocks in the invalid region.

\subsubsection{Reverse Infilling}

The re-synchronization from invalid instructions in the invalid region to valid instructions in the valid region may be attributed to the self-repairing feature of the disassembly process mentioned earlier.
This also suggests that there might be a hidden instruction sequence ending at the start address of the valid region. The reverse infilling process aims to find this sequence.

To find all possible instructions that end at a particular address $a$, we must decode all possible addresses from $a-n$ to $a-1$, where $n=15$ is the maximum instruction length in the x86 architecture, and keep only the instructions that end at $a$. Addresses that have already been identified as invalid instructions are not included.
The process can be continued recursively on the start addresses of these newly found instructions. The results can be considered as a tree, where the root node is $a$, and each child node represents an address decoded into an instruction ending at the address in its parent node. We denote this process as reverse disassembly.

Therefore, given the start address of a valid region, \sysname{} conducts a BFS on this tree to obtain a maximum of $N$ instructions, where $N=16$ is the maximum number of instructions identified at once in the prefilter step.
\sysname{} then obtains the related instructions for these $N$ instructions using the previously mentioned method, gets the probability of each instruction being valid, and retains only those with $p > 0.95$, similar to the process in the prefilter step.
By identifying a valid instruction sequence before the valid region, the start address of the valid region can be extended, and this process is continued recursively at the new start address.

Once no new valid instruction is identified, the reverse infilling switches to single instruction mode. In this mode, instead of performing reverse disassembly recursively, \sysname{} only finds all instructions ending at address $a$, where $a$ is the start address of the latest valid region.
The validity of these instructions is classified similarly to obtain probability values. If none of these values exceed $0.5$, \sysname{} stops the reverse infilling process. Otherwise, it selects the instruction with the highest probability as the valid one and recursively continues the single instruction mode with the new start address of the valid region.

\subsubsection{Forward Infilling}

After completing reverse infilling on an invalid region between two valid ones, \sysname{} begins forward infilling. This step aims to find small valid blocks within the invalid region that are too small to be resynchronized with the number of instructions required by the self-repairing (e.g., containing fewer than three instructions).

To start forward infilling, \sysname{} disassembles one block at the first non-invalid address in the invalid region and identifies the validity of each instruction in the block. The process also works in two stages: a prefilter that confirms the validity of highly certain instructions and a single instruction checker that examines the remaining instructions one by one, ensuring all instructions are identified.
If any valid instructions are detected, the process continues from the end of the last valid instruction; otherwise, it continues from the next non-invalid address. The process stops when the end of the invalid region is reached.

\subsubsection{Example}

To illustrate the effect of disassembly fixing, we use a simplified example snippet extracted from a real obfuscated binary, as shown in Listing \ref{FixingExample}. On the left, we present the initial disassembly results that have already been checked using the method described in \cref{CheckDisasm}.
The instructions in the green frame are identified as valid, designating the corresponding area as a valid region. Conversely, the red frame marks the invalid region containing invalid instructions. Since the invalid region lies between two valid regions, \sysname{} attempts to fix the disassembly within the invalid region.
The right part illustrates the same regions after the instructions in the invalid region have been fixed, assuming the classifier model functions correctly. The lower blue rectangle marks the instruction sequence obtained from reverse infilling, while the upper blue rectangle marks the short block recovered from forward infilling.

\begin{listing}[!ht]
\begin{minipage}{0.23\textwidth}
\begin{mdframed}[skipabove=0pt, skipbelow=0pt, innerleftmargin=1pt, innerrightmargin=1pt, innertopmargin=1pt, innerbottommargin=1pt, linecolor=green]
\begin{minted}[fontsize=\fontsize{7pt}{9pt}, escapeinside=||]{nasm}
mov DWORD PTR [rbp-0x8], eax
mov DWORD PTR [rbp-0x4], 0x8
jmp 0x401007
\end{minted}
\end{mdframed}
\begin{mdframed}[skipabove=0pt, skipbelow=0pt, innerleftmargin=1pt, innerrightmargin=1pt, innertopmargin=1pt, innerbottommargin=1pt, linecolor=red]
\begin{minted}[fontsize=\fontsize{7pt}{9pt}, escapeinside=||]{nasm}
mov edi, eax
rex.RB cld
(bad)
add BYTE PTR [rax], al
add bl, ch
(bad)
lea edi, ds:0x402000
\end{minted}
\end{mdframed}
\begin{mdframed}[skipabove=0pt, skipbelow=0pt, innerleftmargin=1pt, innerrightmargin=1pt, innertopmargin=1pt, innerbottommargin=1pt, linecolor=green]
\begin{minted}[fontsize=\fontsize{7pt}{9pt}, escapeinside=||]{nasm}
mov al, 0x0
call 0x401000
\end{minted}
\end{mdframed}
\end{minipage}%
\hspace{0.01\textwidth}\vline\hspace{2pt}%
\begin{minipage}{0.237\textwidth}
\begin{mdframed}[skipabove=0pt, skipbelow=0pt, innerleftmargin=1pt, innerrightmargin=1pt, innertopmargin=1pt, innerbottommargin=1pt, linecolor=green]
\begin{minted}[fontsize=\fontsize{7pt}{9pt}, escapeinside=||]{nasm}
mov DWORD PTR [rbp-0x8], eax
mov DWORD PTR [rbp-0x4], 0x8
jmp 0x401007
\end{minted}
\end{mdframed}
\begin{minted}[fontsize=\fontsize{7pt}{9pt}, escapeinside=||]{nasm}
db 0x89
\end{minted}
\begin{mdframed}[skipabove=0pt, skipbelow=0pt, innerleftmargin=1pt, innerrightmargin=1pt, innertopmargin=1pt, innerbottommargin=1pt, linecolor=blue]
\begin{minted}[fontsize=\fontsize{7pt}{9pt}, escapeinside=||]{nasm}
mov DWORD PTR [rbp-0x4], 0x17
jmp 0x401007
\end{minted}
\end{mdframed}
\begin{minted}[fontsize=\fontsize{7pt}{9pt}, escapeinside=||]{nasm}
db 0xA9
\end{minted}
\begin{mdframed}[skipabove=0pt, skipbelow=0pt, innerleftmargin=1pt, innerrightmargin=1pt, innertopmargin=1pt, innerbottommargin=1pt, linecolor=blue]
\begin{minted}[fontsize=\fontsize{7pt}{9pt}, escapeinside=||]{nasm}
lea rdi, ds:0x402000
\end{minted}
\end{mdframed}
\begin{mdframed}[skipabove=0pt, skipbelow=0pt, innerleftmargin=1pt, innerrightmargin=1pt, innertopmargin=1pt, innerbottommargin=1pt, linecolor=green]
\begin{minted}[fontsize=\fontsize{7pt}{9pt}, escapeinside=||]{nasm}
mov al, 0x0
call 0x401000
\end{minted}
\end{mdframed}
\end{minipage}
\caption{An example of disassembly fixing.}
\label{FixingExample}
\end{listing}

\subsection{Batching} \label{Batching}

To further improve the efficiency of \sysname{}, we exploit the parallel capabilities of GPUs by using the batching mechanism supported by the transformer library~\cite{Wolf2019HuggingFacesTS}.
For each of the processes described above, classification tasks are not executed immediately. Instead, they are pushed into a queue with the data required to post-process the results after classification.
When the number of items in the queue reaches $M$, the maximum batch size, \sysname{} pops out the batch for classification and then post-processes the result probabilities using the corresponding data.

\subsection{Representing Blocks of Instructions} \label{RepresentInst}

Given (split) blocks extracted from the disassembly graph using the previously mentioned method, we must represent them as text to leverage LLM for the code understanding task.
The blocks should first be sorted in ascending order according to their start addresses so that when presented as text, they also follow the same order.
Representing a block is a straightforward process of displaying all its instructions in Intel syntax. Additionally, for each instruction with extra information, a comment is appended after the instruction. For example, if the instruction is already known to be valid or invalid, a comment \texttt{; valid} or \texttt{; invalid} can be appended.
However, unlike assembly code typically written by programmers, these blocks have some special features that need to be tackled.

\subsubsection{Handling Gap} \label{HandleGap}

The extracted blocks can be noncontinuous; in other words, in the sorted list of blocks $B$, the start address of block $B[i+1]$ is larger than that of its preceding block $B[i]$. In such cases, we must indicate this gap in the text to inform the LLM.
To achieve this, we mark each continuous region's start and end addresses consisting of one or more adjacent blocks preceding and following the instructions. Examples of the text representation of these regions can be found in Figure \ref{OverlapMinExample}.
The start address is marked like a label, while the end address is marked in a comment. For example, the region \texttt{0x401000-0x401006} consists of two adjacent blocks with ranges \texttt{0x401000-0x401004} and \texttt{0x401004-0x401006}.
Moreover, if a block in the middle of the region is referenced by any instruction in the list of blocks (e.g., \texttt{jmp}), its start address should also be marked.

\subsubsection{Handling Overlap} \label{HandleOverlap}

\begin{listing}[!ht]
\begin{minipage}{0.24\textwidth}
\begin{minted}[fontsize=\fontsize{7pt}{9pt}, escapeinside=||]{nasm}
0x401000:
cmp eax, eax
je 0x401007
; 0x401004

<<<<<<<
0x401004:
nop
int3
call 0x4011c3
add BYTE PTR [rdi+0x1], bh
; 0x401011
=======
0x401007:
mov eax, 0x1
mov edi, 0x1
; 0x401011
>>>>>>>

0x401011:
movabs rsi, 0x402000
mov edx, 0xe
syscall
; 0x401022
\end{minted}
\end{minipage}%
\vline\hspace{2pt}%
\begin{minipage}{0.23\textwidth}
\begin{minted}[fontsize=\fontsize{7pt}{9pt}, escapeinside=||]{nasm}
0x401000:
cmp eax, eax
je 0x401007
nop
int3
; 0x401006

<<<<<<<
0x401006:
call 0x4011c3
add BYTE PTR [rdi+0x1], bh
; 0x401011
=======
0x401007:
mov eax, 0x1
mov edi, 0x1
; 0x401011
>>>>>>>

0x401011:
movabs rsi, 0x402000
mov edx, 0xe
syscall
; 0x401022
\end{minted}
\end{minipage}
\caption{An example of overlap minimization.}
\label{OverlapMinExample}
\end{listing}

Another possible cause is that blocks can overlap with each other: in the sorted list of blocks $B$, the start address of $B[i+1]$ is smaller than the end address of $B[i]$. This can frequently occur if the binary is obfuscated, such as when an opaque predicate causes disassembly desynchronization~\cite{Tung2020AHA} (as shown in Listing \ref{OverlapMinExample}).
To present such overlap more explicitly, we use a special technique to convey the overlap information in the text.
Firstly, \sysname{} groups all overlapping blocks in a transitive relation. In other words, two blocks that do not overlap are still in the same group if they both overlap with another common block. This problem can be considered a variant of the problem of merging overlapping intervals~\cite{MergeOverlap}.
We describe the details of this algorithm in \cref{GroupOverlap}.
The strategy described in \cref{HandleGap} represents groups with a single block. For groups with more than one block (i.e., those with overlaps), \sysname{} wraps them using a special notation similar to the git conflict marker~\cite{GitConflict}.
An example of such representation is shown in the left part of Listing \ref{OverlapMinExample}, where the block with the range \texttt{0x401004-0x401011} and the block with the range \texttt{0x401007-0x401011} overlap with each other and are wrapped with markers.

However, simply using such a representation does not effectively demonstrate the overlap. For example, if block $b_1$ overlaps with block $b_2$, but only a few of the ending instructions in $b_1$ overlap with $b_2$, while $b_1$ is a large block.
In this case, using the conflict marker to wrap the two whole blocks does not convey the overlap in the most representative manner. Instead, we should \textit{minimize the overlap region} as much as possible by splitting the overlapped blocks.
To achieve this, we apply the same algorithm for grouping overlapping intervals described in \cref{GroupOverlap} to all instructions in the overlapped group (i.e., multiple transitive overlapped blocks), treating each instruction as an interval. This results in a list of groups consisting of possibly overlapped instructions.
For all groups consisting of multiple instructions (i.e., where the overlap exists), we split the basic block at both the start and end addresses of the group to minimize the overlapped regions.

\subsection{Group Overlapping Intervals} \label{GroupOverlap}

\vspace{2pt}\noindent\textbf{Problem statement}.
The input of the algorithm is a list of intervals $I = \{ (a_i, b_i) \mid i = 0, 1, 2, \ldots, n-1 \}$, where $n$ is the number of intervals. In our scenario, the interval is the start address and the end address of either an instruction or a block.
We consider two intervals $(a_i, b_i)$ and $(a_j, b_j)$ as overlapped if $a_i \leq a_j < b_i$ or $a_j \leq a_i < b_j$. Our definition of overlap is different from others~\cite{MergeOverlap} in the sense that we do not consider $a_j = b_i$ or $a_i = b_j$ as overlap, because adjacent instructions and blocks are not overlapped ones in our scenario.
The task is to group all overlapped intervals in $I$, where the overlap relation is transitive, so two intervals that do not overlap with each other should be in the same group if they both overlap with another interval.

\vspace{2pt}\noindent\textbf{Algorithm}.
To achieve this, we firstly sort $I$ according to $a_i$ in ascending order, and iterate over the sorted list. We always maintain $b'$, the maximum $b$ value of the last group. If the $a$ of the next interval is smaller than $b'$, we add it to the last group and update $b'$; otherwise, create a new group that only contains this interval. The details are shown in Algorithm \ref{GroupAlgo}.

\begin{algorithm}
\caption{Group Overlapping Intervals} \label{GroupAlgo}
\begin{algorithmic}[1]
\REQUIRE $I = \{ (a_i, b_i) \mid i = 0, 1, 2, \ldots, n-1 \}$
\STATE{sort $I$ according to $a$ in ascending order}
\STATE{$b'$ := $I[0].b$}
\STATE{$g$ := \{$I[0]$\}}
\STATE{$G$ := $\{\}$}
\FOR{$i$ = $1$ \TO $n-1$}
	\IF{$I[i].a < b'$}
		\STATE{$g$.append($I[i]$)}
		\STATE{$b'$ := $max(b', I[i].b)$}
	\ELSE
		\STATE{$G$.append($g$)}
		\STATE{$b'$ := $I[i].b$}
		\STATE{$g$ := \{$I[i]$\}}
	\ENDIF
\ENDFOR
\STATE{$G$.append($g$)}
\RETURN $G$
\end{algorithmic}
\end{algorithm}

\section{The LLM-based Validity Classifier}

In this section, we outline the process of fine-tuning Llama 3~\cite{llama3modelcard} to transform it into an encoder model capable of identifying instruction validity through token classification, utilizing techniques proposed in LLM2Vec~\cite{BehnamGhader2024LLM2VecLL}.
First, we describe the forward propagation process, explaining how the probability outputs for validity are generated from the input text snippet of decoded assembly instructions (\cref{ForwardProp}).
Next, we describe the process of fine-tuning the decoder transformer model to develop an encoder model specialized in identifying instruction validity. This encoder model can compute the embedding vector of an instruction within the context of a snippet (\cref{MNTPTraining}).
Finally, we explain the training process for the last linear classifier layer, which generates the probability values for instruction validity (\cref{TrainClassifier}).

\subsection{Forward Propagation} \label{ForwardProp}

We utilize the forward propagation method proposed by LLM2Vec~\cite{BehnamGhader2024LLM2VecLL}, which is generally the same as the forward propagation used in the encoder model for token classification, with slight variations.

First, given a snippet of decoded assembly text with the instructions to be checked separated (i.e., in the form of ``words''), the text is tokenized using the fast tokenizer~\cite{Moi_HuggingFace_s_Tokenizers_2023}. This allows us to determine which ``word'' each token belongs to.
Additionally, a special token \texttt{<s>} is added at the beginning to mark the start of the sequence.
These $k$ tokens are then input into the transformer model~\cite{Vaswani2017AttentionIA}, so that the last hidden layer outputs $k$ vectors matching each of the input tokens. For a token at index $i$, its embedding vector in the last hidden layer is denoted as $e_i$.

The embedding of an instruction constituted by $n$ tokens at indexes $\{i_0, i_0+1, ..., i_0+n-1\}$ is the mean value of the corresponding output vectors in the last hidden layer, computed as follows:

$$e = \frac{\sum_{i=i_0}^{i_0+n-1} e_{i-1}}{n}$$

It is important to note that $e_{i-1}$ is used instead of $e_i$ here: instead of using the output vector of each token, the output vector of the previous token of each token is used.

Finally, the embedding vector of the instruction is input into a linear classifier with an input size equal to the dimension of $e$ and an output size of $1$. The sigmoid function~\cite{Han1995TheIO} is then applied to the output value to calculate the final probability indicating the validity.

\subsection{MNTP Training} \label{MNTPTraining}

Similar to LLM2Vec, we fine-tuned the model using masked next token prediction (MNTP). However, instead of using texts from English Wikipedia, the model is trained on assembly code gathered from compiled binaries.
To generate a large dataset of binaries, we used AnghaBench~\cite{Silva2021ANGHABENCHAS}. This benchmark consists of a substantial number of individually-compilable functions extracted from open-source repositories. We compiled these functions into binaries at different optimization levels.
Additionally, because normal compilation generally does not insert non-code bytes into the code region, we also compiled the functions with junk byte insertion between basic blocks, using the same obfuscation approach described in \cref{CurrentChallenge}.
The ground truth instruction addresses were also gathered by modifying the source code of the \texttt{clang} assembler~\cite{Lattner2004LLVMAC, LLVM}.

For each generated binary, we disassembled it according to the ground truth addresses to obtain all valid instructions.
Additionally, we decoded at incorrect addresses to obtain invalid instructions, placing these invalid instructions into comments. This included addresses of non-code bytes as well as addresses within the middle of valid instructions.
An example snippet of the assembly text used for training is shown in Listing \ref{MNTPText}. Invalid instructions are decoded at all non-code byte addresses, and one random offset in the middle of each multibyte valid instruction is selected to decode an additional invalid instruction.
We do not select all possible offsets within the valid instructions because this would produce too many invalid instructions, making the dataset excessively large and the instructions unbalanced.
The validity of each decoded instruction is also marked at the end of the line with a comment (i.e., \texttt{; valid} and \texttt{; invalid}). Besides, each address referenced by any instruction in the text is labeled with its hexadecimal address value (e.g., \texttt{0x29f3:}).

\begin{listing}[!ht]
\begin{minted}[fontsize=\fontsize{7pt}{9pt}]{nasm}
.byte 0xc9 ; leave  ; invalid
.byte 0x88 ; mov bl, ch ; invalid
jmp 0x29f3 ; valid
; offset 1: add ch, bl ; invalid
jmp 0x29f3 ; valid
; offset 1: add byte ptr [rax - 0x75], cl ; invalid

0x29f3:
mov rsp, qword ptr [rbp - 0x98] ; valid
; offset 2: movsd dword ptr [rdi], dword ptr [rsi] ; invalid
\end{minted}
\caption{Part of the text in the MNTP dataset.}
\label{MNTPText}
\end{listing}

This form of the dataset aims to enable the model not only to learn to be an encoder but also to become more familiar with both valid and invalid assembly code.
According to Llama's reports on how the pretrained dataset is gathered~\cite{Touvron2023LLaMAOA, Touvron2023Llama2O, Rozire2023CodeLO}, the dataset includes only publicly available code and does not contain a large amount of assembly code generated by the compiler, let alone invalidly decoded instructions in the executable.
Therefore, this fine-tuning aims to address and complement this deficiency.

The training settings are generally the same as those in the original work of LLM2Vec~\cite{BehnamGhader2024LLM2VecLL}: the MNTP fine-tuning is performed using LoRA~\cite{Hu2021LoRALA} with $r=16$ and a masked probability of $0.2$.
Additionally, we increased the total number of training steps from 1,000 to 100,000 to allow for a longer training period. This is necessary because our dataset contains much more tokens than the English Wikipedia dataset in total.

\subsection{Classifier Training} \label{TrainClassifier}

After MNTP fine-tuning, we can compute the embedding of an instruction within a snippet of decoded assembly instructions. Using this embedding, we can then train a linear classifier to identify the validity of the instruction.
We follow the practice of LLM2Vec~\cite{BehnamGhader2024LLM2VecLL} by training only the parameters of the linear classifier while keeping the decoder transformer model frozen.
However, since we only have two labels, one neuron with a sigmoid function~\cite{Han1995TheIO} is sufficient, instead of using softmax with two neurons in the last layer.

The supervised dataset used to train the classifier consists of entries where each text is split into ``words'' with corresponding labels, similar to most datasets used in token classification problems, such as \cite{TjongKimSang2003IntroductionTT}.
In our context, the text is an assembly code snippet, and the assembly instructions with their validity labels are separated as ``words''.
A simplified example of such an entry in the dataset is shown in Listing \ref{SupervisedText}. Each region separated by horizontal lines specifies a ``word'' in the text, with the corresponding label on the right-hand side.
The labels \texttt{0} and \texttt{1} represent valid and invalid instructions, respectively, while the label \texttt{-100} indicates ``words'' that are not considered when computing the loss function.

\begin{listing}[!ht]
\begin{minipage}{0.3\textwidth}
\begin{minted}[fontsize=\footnotesize, escapeinside=||]{nasm}
0x8279c:
xchg esp, eax
\end{minted}
\vspace{-0.85em}\rule{\linewidth}{0.4pt}\vspace{-0.35em}
\begin{minted}[fontsize=\footnotesize, escapeinside=||]{nasm}
sub bh, bl
\end{minted}
\vspace{-0.85em}\rule{\linewidth}{0.4pt}\vspace{-0.35em}
\begin{minted}[fontsize=\footnotesize, escapeinside=||]{nasm}
jnp 0x82768
\end{minted}
\vspace{-0.85em}\rule{\linewidth}{0.4pt}\vspace{-0.35em}
\begin{minted}[fontsize=\footnotesize, escapeinside=||]{nasm}
0x827a1:
\end{minted}
\vspace{-0.85em}\rule{\linewidth}{0.4pt}\vspace{-0.35em}
\begin{minted}[fontsize=\footnotesize, escapeinside=||]{nasm}
and al, 0x24
\end{minted}
\vspace{-0.85em}\rule{\linewidth}{0.4pt}\vspace{-0.35em}
\begin{minted}[fontsize=\footnotesize, escapeinside=||]{nasm}
or byte ptr [rax], al
\end{minted}
\vspace{-0.85em}\rule{\linewidth}{0.4pt}\vspace{-0.35em}
\begin{minted}[fontsize=\footnotesize, escapeinside=||]{nasm}
add byte ptr [rax], al
\end{minted}
\vspace{-0.85em}\rule{\linewidth}{0.4pt}\vspace{-0.35em}
\begin{minted}[fontsize=\footnotesize, escapeinside=||]{nasm}
jmp 0x827e8
\end{minted}
\vspace{-0.85em}\rule{\linewidth}{0.4pt}\vspace{-0.35em}
\begin{minted}[fontsize=\footnotesize, escapeinside=||]{nasm}
0x827aa:
\end{minted}
\vspace{-0.85em}\rule{\linewidth}{0.4pt}\vspace{-0.35em}
\begin{minted}[fontsize=\footnotesize, escapeinside=||]{nasm}
loop 0x827f4
\end{minted}
\vspace{-0.85em}\rule{\linewidth}{0.4pt}\vspace{-0.35em}
\begin{minted}[fontsize=\footnotesize, escapeinside=||]{nasm}
mov eax, dword ptr [rsp + 0x10]
\end{minted}
\vspace{-0.85em}\rule{\linewidth}{0.4pt}\vspace{-0.35em}
\begin{minted}[fontsize=\footnotesize, escapeinside=||]{nasm}
bswap rax
\end{minted}
\vspace{-0.85em}\rule{\linewidth}{0.4pt}\vspace{-0.35em}
\begin{minted}[fontsize=\footnotesize, escapeinside=||]{nasm}
0x827b3:
mov qword ptr [rsp + 0x38], rax
mov rax, qword ptr [rsp + 0x28]
\end{minted}
\end{minipage}%
\vline%
\begin{minipage}{0.18\textwidth}
\vspace{0.43em}
\begin{minted}[fontsize=\footnotesize, escapeinside=||]{nasm}
 |\textcolor{white}{.}|
 -100
\end{minted}
\vspace{-0.85em}\rule{\linewidth}{0.4pt}\vspace{-0.35em}
\begin{minted}[fontsize=\footnotesize, escapeinside=||]{nasm}
 0
\end{minted}
\vspace{-0.85em}\rule{\linewidth}{0.4pt}\vspace{-0.35em}
\begin{minted}[fontsize=\footnotesize, escapeinside=||]{nasm}
 0
\end{minted}
\vspace{-0.85em}\rule{\linewidth}{0.4pt}\vspace{-0.35em}
\begin{minted}[fontsize=\footnotesize, escapeinside=||]{nasm}
 -100
\end{minted}
\vspace{-0.85em}\rule{\linewidth}{0.4pt}\vspace{-0.35em}
\begin{minted}[fontsize=\footnotesize, escapeinside=||]{nasm}
 0
\end{minted}
\vspace{-0.85em}\rule{\linewidth}{0.4pt}\vspace{-0.35em}
\begin{minted}[fontsize=\footnotesize, escapeinside=||]{nasm}
 0
\end{minted}
\vspace{-0.85em}\rule{\linewidth}{0.4pt}\vspace{-0.35em}
\begin{minted}[fontsize=\footnotesize, escapeinside=||]{nasm}
 0
\end{minted}
\vspace{-0.85em}\rule{\linewidth}{0.4pt}\vspace{-0.35em}
\begin{minted}[fontsize=\footnotesize, escapeinside=||]{nasm}
 1
\end{minted}
\vspace{-0.85em}\rule{\linewidth}{0.4pt}\vspace{-0.35em}
\begin{minted}[fontsize=\footnotesize, escapeinside=||]{nasm}
 -100
\end{minted}
\vspace{-0.85em}\rule{\linewidth}{0.4pt}\vspace{-0.35em}
\begin{minted}[fontsize=\footnotesize, escapeinside=||]{nasm}
 0
\end{minted}
\vspace{-0.85em}\rule{\linewidth}{0.4pt}\vspace{-0.35em}
\begin{minted}[fontsize=\footnotesize, escapeinside=||]{nasm}
 0
\end{minted}
\vspace{-0.85em}\rule{\linewidth}{0.4pt}\vspace{-0.35em}
\begin{minted}[fontsize=\footnotesize, escapeinside=||]{nasm}
 1
\end{minted}
\vspace{-0.85em}\rule{\linewidth}{0.4pt}\vspace{-0.35em}
\begin{minted}[fontsize=\footnotesize, escapeinside=||]{nasm}
 |\textcolor{white}{.}|
 -100
 |\textcolor{white}{.}|
\end{minted}
\end{minipage}
\caption{An example snippet in supervised dataset.}
\label{SupervisedText}
\end{listing}

To obtain this dataset, a supervised dataset is gathered to be specialized for the task required by the disassembly strategy of \sysname{}.
To achieve this, we use the same disassembly strategy described in \cref{Disassembly} on the same binaries used in MNTP training, but replace the validity classifier with the ground truth, ensuring it always provides the correct validity.
Each time the classifier is invoked, the queried assembly snippet is split so that each instruction being checked becomes a single word. These checked instructions are labeled with the ground truth validity value: \texttt{0} for invalid and \texttt{1} for valid. For other texts, the label is marked as \texttt{-100} to inform the trainer that the word is not considered in the loss function.
These include not only non-instruction text such as address labels or overlap markers, but also instructions not checked by the classifier (e.g., \texttt{xchg esp, eax} in Listing \ref{SupervisedText}). Such split texts, along with the corresponding labels, are recorded as an entry in the dataset.
In this way, we can construct the most appropriate behavior for the classifier to learn.
When applied to all binaries of AnghaBench, this approach produces an excessive number of dataset rows required by the linear classifier learning. Therefore, we randomly select a subset of the results.

\section{Evaluation}

\vspace{2pt}\noindent\textbf{Implementation}.
We implement \sysname{} in 2,000+ lines of Python code, along with 200+ lines of Python code to gather the training data. We use Capstone~\cite{Capstone} as the underlying disassembler engine that decodes a byte sequence into an instruction.
To implement the disassembly method described in \cref{Disassembly}, we wrap the disassembly graph in a Python class,
exposing an efficient interface for basic block insertion, modification, and deletion.
To implement the batching mechanism (\cref{Batching}), instead of sharing one queue for classification tasks required by different components, \sysname{} only shares the same queue for classification tasks that have same form of post-processing. For example, the classification tasks requested by the prefilter step are in a different queue from those requested by the reverse infilling.
Additionally, \sysname{} uses a Python abstract interface to represent the classifier primitive, which can either be the LLM-based validity classifier used in real disassembly tasks or one that checks the validity according to the ground truth directly, which is used to gather training data as described in \cref{TrainClassifier}.

\vspace{2pt}\noindent\textbf{Settings}.
We ran all of our experiments on an Ubuntu 22.04.4 LTS system equipped with an NVIDIA H100 GPU and an AMD EPYC 9124 16-Core Processor.

\vspace{2pt}\noindent\textbf{Research questions}.
In the evaluation part, we explore the following research questions through experiments.
\begin{itemize}
\item \textbf{RQ1}: How effective is \sysname{} to disassemble obfuscated executables compared to other disassemblers?
\item \textbf{RQ2}: How accurate is the LLM-based validity classifier when it is used by  the different components of \sysname{} described in \cref{Disassembly}?
\item \textbf{RQ3}: How effective is our fine-tuned model for identifying valid instructions when it is used alone?
\item \textbf{RQ4}: How fast is \sysname{} for the disassembly task?
\end{itemize}

\subsection{Comparison with Other Disassemblers (RQ1)} \label{EvalComparison}

\vspace{2pt}\noindent\textbf{Benchmark}.
We use the same set of real-world open-source programs as used by DeepDi~\cite{Yu2022DeepDiLA},
which contains a wide range of targets ranging from cryptography libraries (\texttt{openssl}) to SQL database engines (\texttt{sqlite3}).
Each program is compiled and obfuscated in the same way as described in \cref{CurrentChallenge} (e.g., OLLVM-based~\cite{Junod2015ObfuscatorLLVMS} control flow flattening~\cite{Lszl2009OBFUSCATINGCP} and junk bytes insertion between flattened blocks).
To ensure the insertion of junk bytes,
we add a bogus control flow (i.e., control flow that never executes) from the \texttt{SwitchInst} of the dispatcher block to all inserted junk bytes.
This obfuscation scheme ensures that junk bytes are inserted before almost all basic blocks, while the control flow information is stripped by the flattening.
Therefore, disassemblers' anti-obfuscation ability can be thoroughly tested.
To obtain the groundtruth, we modify the LLVM assembler~\cite{Lattner2004LLVMAC} to collect the starting addresses of all valid instructions.

\vspace{2pt}\noindent\textbf{Baseline disassemblers}.
We firstly choose DeepDi~\cite{Yu2022DeepDiLA} as the state-of-the-art disassembly approach based on machine learning. We also choose Angr~\cite{Shoshitaishvili2016SOKO} because it is the best industrial disassembler according to the measurement of ~\cite{Pang2020SoKAY}. However, we only use its \texttt{CFGFast} feature instead of \texttt{CFGEmulated} that leverages the symbolic execution, because it throws an exception when applied on our obfuscated binaries. We also chose Ddisasm~\cite{FloresMontoya2019DatalogD} as a heuristic-based baseline.
There are also some disassemblers~\cite{Bardin2017BackwardBoundedDT, Tung2020AHA} aiming to tackle obfuscation. We did not include them because the obfuscations they handle is different from the one used in this benchmark. Moreover, according to the maintainer of BINSEC, BB-DSE~\cite{Bardin2017BackwardBoundedDT} is unavailable now. Besides, we did not include XDA~\cite{Pei2020XDAAR} because its fine-tuned disassembler model is not released, while according to the experiment of DeepDi, they demonstrate similar results, so DeepDi alone is already representative enough for the machine learning approaches.

\vspace{2pt}\noindent\textbf{Results}. The results are shown in Table \ref{CompTable}, which show that \sysname{} outperforms other state-of-the-art disassemblers. Angr and Ddisasm are even unable to correctly disassemble most of the instructions. DeepDi, the second-best disassembler, achieves a decent score on decoding all instructions but fails to decode the first instructions following the junk byte regions. Moreover, we find that \sysname{} achieves higher precision than the recall in general, which means some correct instructions are not decoded. We try to explain this results using intermediate metrics in \cref{IntermMetrics}.

\begin{table*}[h]
\caption{The results of comparing \sysname{} to other disassembler counterparts.}
\label{CompTable}
\footnotesize
\centering
\begin{tabular}{|c|c|c|cccccccccc|}
\hline
Disassembler & Type & Metrics & curl & diffutils & GMP & ImageMagick & libmicrohttpd & libtomcrypt & OpenSSL & PuTTy & SQLite & zlib \\ \hline
\multirow{6}{*}{DisasLLM} & \multirow{3}{*}{All} & Precision & \textbf{0.98} & \textbf{0.98} & \textbf{0.98} & \textbf{0.98} & \textbf{0.98} & \textbf{0.99} & \textbf{0.99} & \textbf{0.99} & \textbf{0.99} & \textbf{0.98} \\ \cline{3-13}
 & & Recall  & 0.92 & \textbf{0.94} & \textbf{0.95} & 0.92 & \textbf{0.94} & 0.94 & 0.91 & 0.93 & 0.93 & \textbf{0.94} \\ \cline{3-13}
 & & F1  & \textbf{0.95} & \textbf{0.96} & \textbf{0.96} & \textbf{0.95} & \textbf{0.96} & 0.96 & 0.95 & 0.96 & \textbf{0.96} & \textbf{0.96} \\ \cline{2-13}
 & \multirow{3}{*}{Junk} & Precision  & \textbf{0.94} & \textbf{0.92} & \textbf{0.90} & \textbf{0.89} & \textbf{0.92} & \textbf{0.91} & \textbf{0.94} & \textbf{0.93} & \textbf{0.94} & \textbf{0.92} \\ \cline{3-13}
 & & Recall  & \textbf{0.91} & \textbf{0.89} & \textbf{0.89} & \textbf{0.86} & \textbf{0.91} & \textbf{0.89} & \textbf{0.87} & \textbf{0.89} & \textbf{0.90} & \textbf{0.91} \\ \cline{3-13}
 & & F1  & \textbf{0.92} & \textbf{0.91} & \textbf{0.89} & \textbf{0.87} & \textbf{0.91} & \textbf{0.90} & \textbf{0.90} & \textbf{0.91} & \textbf{0.92} & \textbf{0.92} \\ \hline
\multirow{6}{*}{DeepDi} & \multirow{3}{*}{All} & Precision & 0.95 & 0.96 & 0.96 & 0.96 & 0.96 & 0.97 & 0.96 & 0.97 & 0.96 & 0.96 \\ \cline{3-13}
 & & Recall  & \textbf{0.93} & 0.93 & 0.92 & \textbf{0.94} & 0.93 & \textbf{0.95} & \textbf{0.94} & \textbf{0.96} & \textbf{0.94} & 0.93 \\ \cline{3-13}
 & & F1  & 0.94 & 0.94 & 0.94 & 0.95 & 0.94 & \textbf{0.96} & \textbf{0.95} & \textbf{0.96} & 0.95 & 0.95 \\ \cline{2-13}
 & \multirow{3}{*}{Junk} & Precision  & 0.60 & 0.62 & 0.65 & 0.57 & 0.59 & 0.67 & 0.63 & 0.65 & 0.61 & 0.58 \\ \cline{3-13}
 & & Recall  & 0.51 & 0.51 & 0.48 & 0.52 & 0.47 & 0.52 & 0.55 & 0.58 & 0.52 & 0.47 \\ \cline{3-13}
 & & F1  & 0.55 & 0.56 & 0.55 & 0.54 & 0.52 & 0.59 & 0.59 & 0.61 & 0.56 & 0.52 \\ \hline
\multirow{6}{*}{Angr} & \multirow{3}{*}{All} & Precision & 0.68 & 0.67 & 0.66 & 0.75 & 0.70 & 0.75 & 0.73 & 0.78 & 0.72 & 0.71 \\ \cline{3-13}
 & & Recall  & 0.74 & 0.71 & 0.77 & 0.80 & 0.73 & 0.84 & 0.79 & 0.82 & 0.75 & 0.74 \\ \cline{3-13}
 & & F1  & 0.71 & 0.69 & 0.71 & 0.77 & 0.71 & 0.80 & 0.76 & 0.80 & 0.73 & 0.72 \\ \cline{2-13}
 & \multirow{3}{*}{Junk} & Precision  & 0.05 & 0.05 & 0.04 & 0.04 & 0.05 & 0.04 & 0.07 & 0.10 & 0.06 & 0.05 \\ \cline{3-13}
 & & Recall  & 0.10 & 0.10 & 0.09 & 0.09 & 0.09 & 0.09 & 0.14 & 0.18 & 0.11 & 0.09 \\ \cline{3-13}
 & & F1  & 0.07 & 0.07 & 0.05 & 0.06 & 0.06 & 0.06 & 0.10 & 0.13 & 0.08 & 0.06 \\ \hline
\multirow{6}{*}{DDisasm} & \multirow{3}{*}{All} & Precision & 0.84 & 0.86 & 0.81 & 0.85 & 0.84 & 0.82 & 0.87 & 0.92 & 0.87 & 0.86 \\ \cline{3-13}
 & & Recall  & 0.23 & 0.21 & 0.19 & 0.12 & 0.18 & 0.22 & 0.33 & 0.41 & 0.26 & 0.17 \\ \cline{3-13}
 & & F1  & 0.37 & 0.34 & 0.31 & 0.22 & 0.30 & 0.35 & 0.48 & 0.56 & 0.40 & 0.29 \\ \cline{2-13}
 & \multirow{3}{*}{Junk} & Precision  & 0.25 & 0.29 & 0.17 & 0.27 & 0.26 & 0.16 & 0.31 & 0.42 & 0.32 & 0.31 \\ \cline{3-13}
 & & Recall  & 0.08 & 0.08 & 0.05 & 0.05 & 0.07 & 0.08 & 0.14 & 0.18 & 0.10 & 0.07 \\ \cline{3-13}
 & & F1  & 0.12 & 0.12 & 0.08 & 0.09 & 0.11 & 0.10 & 0.19 & 0.25 & 0.15 & 0.12 \\ \hline

\end{tabular}
\end{table*}

\subsection{Intermediate Metrics of Classifier Model (RQ2)} \label{IntermMetrics}

In this part, we demonstrate the effectiveness of the LLM-based model when it is used by the different components of \sysname{}. We consider three \sysname{} components that utilizes the classifier: disassembly checking, reverse infilling and forward infilling. Each of these components can be further separated into two components: prefilter component, which checks multiple instructions at once, and single component, which checks each instruction one by one (except for reverse infilling, because there can be multiple instructions ending at the same address). Therefore, we have $3 \times 2 = 6$ components in total. The details about these components have been described previously in \cref{CheckDisasm} and \cref{FixDisasm}.

Therefore, for each of the components, when it uses the LLM-based classifier model, we record the addresses of the instructions being checked and the result probability values, along with the annotation of the component that requests the classification. Based on the ground truth valid instruction addresses, we can then use these records to compute the metrics about the validity identification of each of the components. To be specific, we count the number of false positives, false negatives, true positives and true negatives. The results are presented in Table \ref{IntermTable}.

\begin{table*}[h]
\caption{The metrics of LLM-based validity classifier when being used by the components of \sysname{}.}
\label{IntermTable}
\footnotesize
\centering
\begin{tabular}{|cc|c|ccccccccc|}
\hline
\multicolumn{2}{|c|}{Component} & Metrics & curl & diffutils & GMP & ImageMagick & libmicrohttpd & libtomcrypt & PuTTy & SQLite & zlib \\ \hline
\multicolumn{1}{|c|}{\multirow{8}{*}{Checking}} & \multirow{4}{*}{Prefilter} & FP & 835 & 178 & 917 & 5440 & 262 & 544 & 567 & 1251 & 114\\ \cline{3-12}
\multicolumn{1}{|c|}{}                          &                            & FN & 8190 & 968 & 2894 & 36672 & 1596 & 4393 & 7542 & 14358 & 653\\ \cline{3-12}
\multicolumn{1}{|c|}{}                          &                            & TP & 206261 & 29409 & 130508 & 858529 & 51269 & 145085 & 164896 & 341318 & 21703\\ \cline{3-12}
\multicolumn{1}{|c|}{}                          &                            & TN & 250456 & 36194 & 150463 & 800430 & 54954 & 117969 & 140301 & 372861 & 22050\\ \cline{2-12}
\multicolumn{1}{|c|}{}                          & \multirow{4}{*}{Single}    & FP & 3266 & 530 & 2540 & 15554 & 749 & 1897 & 2137 & 4890 & 309\\ \cline{3-12}
\multicolumn{1}{|c|}{}                          &                            & FN & 5138 & 527 & 1659 & 22804 & 895 & 4524 & 4667 & 9335 & 552\\ \cline{3-12}
\multicolumn{1}{|c|}{}                          &                            & TP & 20832 & 2630 & 8345 & 116998 & 5445 & 12656 & 18127 & 40394 & 2098\\ \cline{3-12}
\multicolumn{1}{|c|}{}                          &                            & TN & 2746 & 362 & 1586 & 9934 & 550 & 1295 & 1362 & 3940 & 199\\ \hline
\multicolumn{1}{|c|}{\multirow{8}{*}{Reverse}}  & \multirow{4}{*}{Prefilter} & FP & 502 & 103 & 355 & 1793 & 155 & 246 & 424 & 879 & 65\\ \cline{3-12}
\multicolumn{1}{|c|}{}                          &                            & FN & 634 & 132 & 56 & 1965 & 192 & 98 & 427 & 863 & 34\\ \cline{3-12}
\multicolumn{1}{|c|}{}                          &                            & TP & 41177 & 6101 & 20114 & 120717 & 9175 & 17877 & 20152 & 62408 & 3738\\ \cline{3-12}
\multicolumn{1}{|c|}{}                          &                            & TN & 75268 & 10484 & 36607 & 255438 & 16249 & 36731 & 36490 & 110532 & 6969\\ \cline{2-12}
\multicolumn{1}{|c|}{}                          & \multirow{4}{*}{Single}    & FP & 357 & 52 & 232 & 1380 & 91 & 166 & 246 & 750 & 40\\ \cline{3-12}
\multicolumn{1}{|c|}{}                          &                            & FN & 222 & 67 & 52 & 963 & 58 & 105 & 164 & 403 & 12\\ \cline{3-12}
\multicolumn{1}{|c|}{}                          &                            & TP & 2423 & 350 & 713 & 14648 & 596 & 2540 & 1710 & 4003 & 194\\ \cline{3-12}
\multicolumn{1}{|c|}{}                          &                            & TN & 586 & 72 & 287 & 2896 & 112 & 284 & 413 & 1024 & 52\\ \hline
\multicolumn{1}{|c|}{\multirow{8}{*}{Forward}}  & \multirow{4}{*}{Prefilter} & FP & 4165 & 597 & 1367 & 10025 & 928 & 1528 & 3345 & 7341 & 381\\ \cline{3-12}
\multicolumn{1}{|c|}{}                          &                            & FN & 7541 & 719 & 1357 & 24308 & 1062 & 2006 & 3602 & 10357 & 327\\ \cline{3-12}
\multicolumn{1}{|c|}{}                          &                            & TP & 20796 & 3073 & 14380 & 61932 & 4697 & 8103 & 9632 & 29355 & 1889\\ \cline{3-12}
\multicolumn{1}{|c|}{}                          &                            & TN & 236686 & 30890 & 149741 & 836289 & 49908 & 111068 & 120236 & 335712 & 19462\\ \cline{2-12}
\multicolumn{1}{|c|}{}                          & \multirow{4}{*}{Single}    & FP & 514 & 75 & 330 & 1840 & 108 & 285 & 357 & 774 & 45\\ \cline{3-12}
\multicolumn{1}{|c|}{}                          &                            & FN & 926 & 164 & 226 & 3015 & 163 & 483 & 610 & 1932 & 63\\ \cline{3-12}
\multicolumn{1}{|c|}{}                          &                            & TP & 1131 & 227 & 293 & 3764 & 198 & 465 & 650 & 2035 & 71\\ \cline{3-12}
\multicolumn{1}{|c|}{}                          &                            & TN & 1551 & 209 & 811 & 5166 & 318 & 731 & 897 & 2384 & 145\\ \hline

\end{tabular}
\end{table*}

Firstly, we can find that during the process of checking the decoded instructions in the initial disassembly results, the prefilter step is already capable of identifying most of the instructions, with a very high accuracy. By contrast, the remaining instructions that are then checked one by one are harder to correctly classify, which aligns with our assumptions. Additionally, false negatives are generally higher than the false positives, especially in the prefilter step. We believe this also explains the metrics in Table \ref{CompTable} where the recall values of \sysname{} are lower than the precision values.

During the process of fixing the disassembly, the scenarios are almost similar for both reverse infilling and forward infilling: the prefilter (i.e., checking multiple decoded instructions as fixing candidates at once) handles most of the instructions correctly, while the single-instruction step performs worse. Moreover, there are many true negatives in the prefilter of forward infilling, because most of the decoded instructions in the invalid regions are invalid ones, and they are successfully identified when multiple instructions are checked simultaneously.

These intermediate results suggest that the model indeed works with high accuracy when used by the different components of \sysname{}.
However, since there are still considerable false positives and false negatives, the model capability can be further improved by better fine-tuning.

\subsection{Capability of Fine-tuned Model (RQ3)} \label{EvalModel}

In this part, we conduct an ablation study investigating the capability of our fine-tuned model when it is used alone, and compare its capability with some other LLM baselines. To be specific, we utilize a naive approach for disassembly: we use LLM model to check the validity of instruction decoded at each address in the code region, and output all addresses that are considered as valid.

However, LLM does not obtain enough information if we only provide the decoded instruction at the address to be checked. Therefore, we also provide the surrounding instructions of start addresses within the range $a-50$ and $a+50$, where $a$ is the address to be checked. We utilize two approaches to provide the surrounding instructions: 1) We decode all addresses within the range (i.e., superset disassembly) and provide all of them to LLM. 2) We conduct linear disassembly starting from $a-50$ and $a$ to get the instructions from both, and show the possible overlap using the format we mentioned in \cref{HandleOverlap}. Note that these two approaches both ensure the instruction to be checked at $a$ is decoded.

\begin{table*}[h]
\caption{Execution speed of \sysname{}.}
\label{SpeedTable}
\footnotesize
\centering
\begin{tabular}{|c|cccccccccc|}
\hline
 & curl & diffutils & GMP & ImageMagick & libmicrohttpd & libtomcrypt & OpenSSL & PuTTy & SQLite & zlib \\ \hline
Time (seconds) & 11588.09 & 1595.89 & 6450.46 & 44274.76 & 2490.57 & 5625.90 & 75514.68 & 6508.85 & 18290.14 & 990.12\\ \hline
Size (bytes) & 1616997 & 227299 & 1040736 & 6605067 & 372941 & 1052911 & 8343974 & 1118449 & 2531104 & 155647\\ \hline
Speed (bytes/second) & 139.54 & 142.43 & 161.34 & 149.18 & 149.74 & 187.15 & 110.49 & 171.84 & 138.39 & 157.20\\ \hline

\end{tabular}
\end{table*}

To enable decoder LLM model to conduct such task, we leveraged the prompt engineering. In the prompt, we introduce the task of identifying the validity of the instruction by informing LLM how instruction can be invalid, the format of the provided assembly snippet, and how the snippet is disassembled (i.e., superset or linear). Finally, we provide the disassembled snippet and ask if a particular instruction is valid, and the response is limited to generate only one character: `Y' for valid or `N' for invalid.

We chose Codestral 22B~\cite{Codestral} and Llama3 8B~\cite{llama3modelcard} as the baseline models for this task, because Codestral is the current state-of-the-art open-weight code model, and Llama3 8B is the model that \sysname{} is fine-tuned on. We also tested Meta LLM Compiler~\cite{Cummins2024MetaLL}, the state-of-the-art model that has been trained on assembly text. However, the model fails to follow our instruction of answering only `Y' or `N', so we exclude it from our results. We did not include any online proprietary models, since the task requires so many tokens that the costs would be too high. Additionally, we also include DeepDi, because it also provides API that identifies if an offset is the start of a valid instruction given the byte sequence. Note that the DeepDi here works differently than the DeepDi in Table \cref{CompTable}, where it checks all addresses at once, while the DeepDi only checks one address for each classification.
For fine-tuned model of \sysname{}, we only tested the linear disassembly, because our model is not trained on the dataset of superset disassembly, which clearly cannot give decent results. Since checking all possible addresses can be too slow for LLM, we only consider the first page of the code region (i.e., 4096 bytes) of each obfuscated binary.

The results are presented in Table \cref{AblationTable} with the same format as Table \cref{CompTable}. We mark the subscript $L$ for linear disassembly, and subscript $S$ for superset disassembly.
Firstly, we can find that \sysname{}- where only the LLM-based validity classifier alone is used to disassemble the binary, produces worse results than the whole system of \sysname{} presneted in Table \cref{CompTable}. This shows that our disassembly strategy is not only more efficient but also more effective. However, our model is still better than DeepDi, whose results are not very different from those in \cref{CompTable}.
Additionally, we can find that directly applying the decoder LLM on such classification task does not yield good results: they almost identify all instructions as valid, so precision is too low while the recall is pretty high. Besides, leveraging superset disassembly does not generate better results than the linear disassembly, which suggests that the superset disassembly indeed provides too much redundant information that is not required by the validity identification.

\begin{table*}[h]
\caption{The results of using models directly on the disassembly.}
\label{AblationTable}
\footnotesize
\centering
\begin{tabular}{|c|c|c|cccccccccc|}
\hline
Disassembler & Type & Metrics & curl & diffutils & GMP & ImageMagick & libmicrohttpd & libtomcrypt & OpenSSL & PuTTy & SQLite & zlib \\ \hline
\multirow{6}{*}{\sysname{}-} & \multirow{3}{*}{All} & Precision & 0.90 & 0.88 & 0.88 & 0.93 & 0.93 & 0.92 & 0.92 & 0.91 & 0.92 & 0.92 \\ \cline{3-13}
 & & Recall  & 0.92 & 0.91 & 0.91 & 0.89 & 0.94 & 0.93 & 0.87 & 0.91 & 0.94 & 0.97 \\ \cline{3-13}
 & & F1  & 0.91 & 0.90 & 0.89 & 0.91 & \textbf{0.94} & 0.92 & 0.89 & 0.91 & 0.93 & 0.94 \\ \cline{2-13}
 & \multirow{3}{*}{Junk} & Precision  & \textbf{0.83} & \textbf{0.75} & \textbf{0.85} & \textbf{0.87} & \textbf{0.92} & \textbf{0.79} & \textbf{0.81} & \textbf{0.86} & \textbf{0.90} & \textbf{0.83} \\ \cline{3-13}
 & & Recall  & \textbf{1.00} & 0.92 & \textbf{1.00} & 0.95 & 0.94 & 0.95 & 0.87 & 0.92 & 0.95 & 0.96 \\ \cline{3-13}
 & & F1  & \textbf{0.91} & \textbf{0.83} & \textbf{0.92} & \textbf{0.91} & \textbf{0.93} & \textbf{0.86} & \textbf{0.84} & \textbf{0.89} & \textbf{0.93} & \textbf{0.89} \\ \hline
\multirow{6}{*}{DeepDi} & \multirow{3}{*}{All} & Precision & \textbf{0.94} & \textbf{0.97} & \textbf{0.96} & \textbf{0.96} & \textbf{0.96} & \textbf{0.98} & \textbf{0.96} & \textbf{0.94} & \textbf{0.96} & \textbf{0.97} \\ \cline{3-13}
 & & Recall  & 0.91 & 0.91 & 0.95 & 0.95 & 0.91 & 0.93 & 0.94 & 0.95 & 0.97 & 0.97 \\ \cline{3-13}
 & & F1  & \textbf{0.93} & \textbf{0.94} & \textbf{0.96} & \textbf{0.95} & 0.93 & \textbf{0.95} & \textbf{0.95} & \textbf{0.95} & \textbf{0.97} & \textbf{0.97} \\ \cline{2-13}
 & \multirow{3}{*}{Junk} & Precision  & 0.55 & 0.68 & 0.68 & 0.66 & 0.60 & 0.65 & 0.68 & 0.59 & 0.73 & 0.63 \\ \cline{3-13}
 & & Recall  & 0.49 & 0.48 & 0.66 & 0.63 & 0.45 & 0.44 & 0.61 & 0.64 & 0.79 & 0.61 \\ \cline{3-13}
 & & F1  & 0.52 & 0.56 & 0.67 & 0.64 & 0.52 & 0.53 & 0.64 & 0.61 & 0.76 & 0.62 \\ \hline
\multirow{6}{*}{Codestral$_S$} & \multirow{3}{*}{All} & Precision & 0.28 & 0.25 & 0.25 & 0.25 & 0.24 & 0.22 & 0.27 & 0.27 & 0.29 & 0.25 \\ \cline{3-13}
 & & Recall  & 0.98 & 0.99 & \textbf{0.99} & \textbf{0.99} & 0.99 & \textbf{1.00} & \textbf{1.00} & \textbf{1.00} & \textbf{0.99} & 0.99 \\ \cline{3-13}
 & & F1  & 0.44 & 0.39 & 0.40 & 0.40 & 0.39 & 0.36 & 0.43 & 0.42 & 0.45 & 0.40 \\ \cline{2-13}
 & \multirow{3}{*}{Junk} & Precision  & 0.22 & 0.18 & 0.19 & 0.21 & 0.18 & 0.15 & 0.21 & 0.22 & 0.22 & 0.18 \\ \cline{3-13}
 & & Recall  & 1.00 & \textbf{1.00} & 1.00 & \textbf{1.00} & \textbf{1.00} & \textbf{1.00} & \textbf{1.00} & \textbf{1.00} & \textbf{1.00} & \textbf{1.00} \\ \cline{3-13}
 & & F1  & 0.36 & 0.31 & 0.31 & 0.35 & 0.31 & 0.26 & 0.35 & 0.37 & 0.36 & 0.31 \\ \hline
\multirow{6}{*}{Codestral$_L$} & \multirow{3}{*}{All} & Precision & 0.28 & 0.26 & 0.25 & 0.29 & 0.29 & 0.26 & 0.28 & 0.28 & 0.32 & 0.31 \\ \cline{3-13}
 & & Recall  & \textbf{0.99} & \textbf{1.00} & 0.99 & 0.99 & \textbf{1.00} & 1.00 & 0.99 & 1.00 & 0.98 & \textbf{1.00} \\ \cline{3-13}
 & & F1  & 0.43 & 0.41 & 0.39 & 0.45 & 0.45 & 0.41 & 0.44 & 0.44 & 0.49 & 0.47 \\ \cline{2-13}
 & \multirow{3}{*}{Junk} & Precision  & 0.19 & 0.18 & 0.18 & 0.22 & 0.20 & 0.14 & 0.21 & 0.20 & 0.24 & 0.21 \\ \cline{3-13}
 & & Recall  & 1.00 & 1.00 & 0.99 & 0.99 & 0.99 & 0.99 & 1.00 & 0.99 & 0.99 & 1.00 \\ \cline{3-13}
 & & F1  & 0.32 & 0.31 & 0.30 & 0.37 & 0.33 & 0.25 & 0.35 & 0.33 & 0.39 & 0.35 \\ \hline
\multirow{6}{*}{Llama$_S$} & \multirow{3}{*}{All} & Precision & 0.30 & 0.27 & 0.26 & 0.32 & 0.33 & 0.29 & 0.33 & 0.31 & 0.34 & 0.34 \\ \cline{3-13}
 & & Recall  & 0.92 & 0.95 & 0.93 & 0.85 & 0.90 & 0.94 & 0.85 & 0.90 & 0.85 & 0.93 \\ \cline{3-13}
 & & F1  & 0.45 & 0.43 & 0.41 & 0.47 & 0.49 & 0.44 & 0.47 & 0.46 & 0.49 & 0.49 \\ \cline{2-13}
 & \multirow{3}{*}{Junk} & Precision  & 0.24 & 0.25 & 0.21 & 0.28 & 0.27 & 0.18 & 0.30 & 0.27 & 0.34 & 0.27 \\ \cline{3-13}
 & & Recall  & 0.90 & 0.95 & 0.89 & 0.81 & 0.86 & 0.91 & 0.87 & 0.87 & 0.79 & 0.93 \\ \cline{3-13}
 & & F1  & 0.38 & 0.39 & 0.33 & 0.42 & 0.42 & 0.30 & 0.44 & 0.41 & 0.48 & 0.42 \\ \hline
\multirow{6}{*}{Llama$_L$} & \multirow{3}{*}{All} & Precision & 0.27 & 0.25 & 0.24 & 0.28 & 0.29 & 0.25 & 0.27 & 0.27 & 0.31 & 0.30 \\ \cline{3-13}
 & & Recall  & 0.97 & 0.99 & 0.97 & 0.93 & 0.98 & 0.93 & 0.94 & 0.97 & 0.94 & 0.97 \\ \cline{3-13}
 & & F1  & 0.42 & 0.40 & 0.39 & 0.43 & 0.45 & 0.39 & 0.42 & 0.42 & 0.46 & 0.45 \\ \cline{2-13}
 & \multirow{3}{*}{Junk} & Precision  & 0.20 & 0.20 & 0.18 & 0.23 & 0.21 & 0.15 & 0.21 & 0.20 & 0.25 & 0.22 \\ \cline{3-13}
 & & Recall  & 0.96 & 0.99 & 0.96 & 0.94 & 0.99 & 0.94 & 0.96 & 0.96 & 0.95 & 0.97 \\ \cline{3-13}
 & & F1  & 0.33 & 0.33 & 0.30 & 0.37 & 0.35 & 0.26 & 0.35 & 0.32 & 0.39 & 0.36 \\ \hline

\end{tabular}
\end{table*}

\subsection{Efficiency of Disassembly (RQ4)}

Finally, we measure the efficiency of \sysname{}. We record the execution time of \sysname{} on each executable generated in \cref{EvalComparison} and calculate the average processing speed in ``bytes per second''. The batch size $M$ was set to $32$ in this experiment. The results are presented in Table \ref{SpeedTable}.

According to the results, \sysname{} disassembles the binary in roughly 100+ bytes per second,
meaning that an executable containing millions of bytes (MBs in size) will need hours to be disassembled.
Although this is slower than some other disassemblers like DeepDi (while still faster than expensive approaches like symbolic execution-based ones), 
\sysname{} has the advantage of effectively handling the obfuscation and producing more accurate results.
We thus consider it a valuable and practical tradeoff.
We leave the performance optimization as a future work.

\section{Discussion}

\vspace{2pt}\noindent\textbf{Efficiency optimization}.
Although already optimized using prefiltering and batching, when disassembling large binary with millions of bytes, \sysname{} still needs to take many hours to accomplish the task. To address this issue, one may suggest not to check all instructions, but instead, if one instruction is identified as valid, the instructions that must follow its control flow should also be valid. However, the problem of this claim is that the inserted junk bytes can appear like valid instructions: while decoding them does not affect the disassembly results since they do not overlap with the real valid instructions, the instructions following their control flow may cause such overlap. Thus, it is still necessary to check all instructions.
Another potential approach is to use a smaller model, such as CodeBERT~\cite{Feng2020CodeBERTAP}, for prefiltering, so that validity of each instruction that is obvious even from the perspective of a small language model can be confirmed first. We leave this direction as the potential future work.

\vspace{2pt}\noindent\textbf{Valid overlapping instructions}.
In general, even in the obfuscated binary, two overlapping instructions cannot be both valid. An exception case, also handled by CoDisasm~\cite{Bonfante2015CoDisasmMS}, involves a \texttt{jmp} instruction jumping into itself.
For example, consider the disassembly of the byte sequence \texttt{eb ff c9}, where \texttt{eb ff} is decoded to \texttt{jmp +1} with target address pointing to \texttt{ff c9}, which is decoded to \texttt{dec ecx}. Since the two overlapping instructions are both executed at runtime, they are both valid.
In principle, \sysname{} is capable of handling this case if such pattern occurs in the initial disassembly graph, as long as the model has learned this pattern, because our disassembly method does not intentionally resolve all overlapped blocks. However, if such pattern occurs in the invalid region which is going to be fixed, \sysname{} is not able to disassemble this pattern correctly because we assume no overlap in the invalid region when it is being fixed.

Another more extreme case involves hiding all real instructions into immediate numbers of another set of valid instructions~\cite{Jamthagen2013ANI}. Since such obfuscation technique aims to defeat the self-repairing feature of the disassembly, current design of \sysname{} cannot handle this scenario. A potential future extension that may allow \sysname{} to handle this involves training another classifier to identify if immediate number of a valid instruction may hide another instruction, and if the classifier says yes, \sysname{} can start an extra disassembly starting from the immediate number of the instruction and check if they are valid.

\vspace{2pt}\noindent\textbf{Combining with traditional disassembly techniques}.
Currently, our initial disassembly process only leverages the most basic traditional disassembly approach, which does not include resolving target addresses of an indirect branch, currently handled by symbolic execution~\cite{Shoshitaishvili2016SOKO} or heuristic-based approach~\cite{Pang2020SoKAY}. These techniques can also be applied upon our initial disassembly algorithm to get a more comprehensive initial disassembly graph.

\section{Related Work}

\subsection{Disassembly and Deobfuscation}

\vspace{2pt}\noindent\textbf{Disassembly of normal binary}.
Commonly-used industrial disassemblers include IDA~\cite{IDAPro}, Ghidra~\cite{Ghidra}, Angr~\cite{Shoshitaishvili2016SOKO}, Binary Ninja~\cite{BinaryNinja}, Radare2~\cite{Radare2}, BAP~\cite{BAP} and objdump~\cite{Objdump}. According to investigations of previous works~\cite{Pang2020SoKAY, Andriesse2016AnIA}, these disassemblers generally leverage linear or/and recursive disassembly, along with additional algorithms and heuristics to improve the effectiveness. \sysname{} is similar to them in the sense that the algorithm is our LLM-based method.
Datalog disassembly~\cite{FloresMontoya2019DatalogD} leverages datalog, a declarative logic programming language, to implement the algorithm and heuristics for binary disassembly, such as the heuristic to resolve overlapping blocks.
Probabilistic disassembly~\cite{Miller2019ProbabilisticD} computes probability for each instruction to be valid based on some heuristics known as hints. By contrast, \sysname{} also computes such probability but \sysname{} relies on LLM-based approach instead of any heuristic.
The disassembler of SAFER~\cite{Priyadarshan2023AccurateDO} also leverages a probability-based approach combined with static analysis using a prioritized error correction algorithm.
XDA~\cite{Pei2020XDAAR} pre-trained an encoder transformer with dataset of byte sequences in the compiled binaries, using masked token prediction as the training objective, which was then fine-tuned for instruction address identification. \sysname{} also leverages a similar procedure, but it was trained on decoded instructions instead of raw bytes, which contain clearer semantic information for the model to learn.
DeepDi~\cite{Yu2022DeepDiLA} disassembles all possible addresses of the code region to construct an instruction flow graph for identifying the validity of each instruction using a graph neuron network. However, as we have demonstrated, its resilience to obfuscated code is limited. %

\vspace{2pt}\noindent\textbf{Disassembly of obfuscated binary}.
Even if disassembling normal binary is already a challenging task, some previous works also aimed to disassemble obfuscated binary.
An early work~\cite{Krgel2004StaticDO} proposes a method to statically disassemble obfuscated binary, which mainly relies on statistical and heuristic methods to exclude invalid instructions and find valid instructions.
Another early work~\cite{Krishnamoorthy2009StaticDO} leverages the decision tree, through feature engineering on the decoded instructions, to detect invalid disassembly caused by the obfuscation.
The opaque predicates detective~\cite{Tung2020AHA} applies a series of heuristics to identify the unreachable branch of an opaque predicate that causes desynchronization of the disassembly.
CoDisasm~\cite{Bonfante2015CoDisasmMS} aims to extract and disassemble self-modifying obfuscated binary. Their disassembly approach relies on the dynamic trace and is resilient to obfuscation technique containing overlapping instructions that are both valid.
Backward-Bounded DSE~\cite{Bardin2017BackwardBoundedDT} utilizes the dynamic trace for concolic execution in order to decide infeasible paths, which provide complementary information for their sparse disassembly method aiming to be resilient to opaque predicate and call stack tampering.
However, these two works above both heavily depend on dynamic trace, so uncovered paths that in general constitute most parts of the program cannot benefit much from these methods, because achieving a high program coverage is a problem known to be hard. \sysname{}, on the other hand, not only performs well without any execution trace, but also can benefit from execution trace by providing them as priorly known valid instruction addresses.

\vspace{2pt}\noindent\textbf{Other deobfuscation techniques}.
Besides defeating obfuscation causing disassembly desynchronization, other obfuscation techniques and approaches to deobfuscate them also exist.
A generic deobfuscation approach~\cite{Yadegari2015AGA} leverages taint propagation and dependency analysis on the execution trace of an obfuscated executable to pinpoint instructions with semantic significance.
Another work~\cite{Yadegari2015SymbolicEO} proposes methods to improve the effectiveness of symbolic execution on the obfuscated code.
LOOP~\cite{Ming2015LOOPLO} also utilizes concolic execution on the execution trace to identify the opaque predicates in the trace.
All of these works similarly require a (symbolic) execution trace of the obfuscated program, unlike \sysname{}.

\subsection{LLM for Program Analysis}

Recent advancement of the large language model has contributed significantly to the field of program analysis, including fuzzing~\cite{Meng2024LargeLM, Eom2024CovRLFJ, Ackerman2023LargeLM, Xia2023Fuzz4AllUF}, static bug detection~\cite{Chen2024WitheredLeafFE, Li2024EnhancingSA, Sun2024LLM4VulnAU, Sun2023GPTScanDL, Wei2024DemystifyingAD} and code repair~\cite{Xia2023AutomatedPR, Joshi2022RepairIN, Fan2022AutomatedRO, Pearce2021ExaminingZV, Jin2023InferFixEP}.

\vspace{2pt}\noindent\textbf{Application on reverse engineering}.
A pioneering measurement study~\cite{Pearce2022PopQC}, by prompting to ask LLM questions about the decompiled code, firstly investigates the ability of LLM to help reverse engineering. %
DeGPT~\cite{Hu2024DeGPTOD} follows such direction by designing a prompt system to optimize the output from the decompiler, along with a checker that ensures the optimized output from LLM has the same semantic as the original decompiled code.
LLM4Decompile~\cite{Tan2024LLM4DecompileDB} fine-tunes the LLM to aid the decompilation task. The model supports both decompilation from assembly to C code directly and refinement of the decompiled code from Ghidra to produce better decompiled code.
Nova~\cite{Jiang2023NovaGL} utilizes hierarchical attention and contrastive learning to improve LLM's capability on understanding and generating the assembly code.
BinaryAI~\cite{Jiang2024BinaryAIBS} leverages LLM for the task of software composition analysis by computing the embeddings of functions in the binary and matching them with source functions of the third-party libraries.
LATTE~\cite{Liu2023HarnessingTP} improves the binary taint analysis with LLM by automating the manual components of taint analysis, using static analysis and prompt engineering.
Although similar to \sysname{}, these works all try to improve reverse engineering by the usage of LLM, they contribute to the different aspects of the reverse engineering. By contrast, to the best of our knowledge, \sysname{} is the first work that leverages LLM for the disassembly task.

\section{Conclusion}

In this work, we proposed \sysname{}, an end-to-end system for disassembling obfuscated binary executables by effectively leveraging the capability of large language model. \sysname{} consists of a fine-tuned LLM-based validity classifier aiming to identify if decoded instructions are correct, and a disassembly strategy that utilizes such classifier to disassemble the obfuscated executable in an end-to-end manner. According to our evaluation, the approach is effective on the disassembly task compared to other state-of-the-art disassemblers, showing the potential of our method. We believe that our research has opened a new direction in leveraging LLM on the disassembly task.

\bibliographystyle{IEEEtranS}
\bibliography{IEEEabrv,ref.bib}



\end{document}